\documentclass[%
 reprint,
 amsmath,amssymb,
 aps,
]{revtex4-1}

\usepackage{graphicx}
\usepackage{dcolumn}
\usepackage{bm}

\usepackage{pst-all}
\input epsf
\usepackage{epic}
\usepackage{pstricks}
\usepackage{epsfig}
\usepackage{pst-grad} 
\usepackage{pst-plot} 
\usepackage[space]{grffile} 
\usepackage{etoolbox} 
\makeatletter 
\patchcmd\Gread@eps{\@inputcheck#1 }{\@inputcheck"#1"\relax}{}{}
\makeatother

\def\be{\begin{equation}}
\def\ee{\end{equation}}

\def\bx{\bm x}

\def\bn{\bm n}
\def\bN{\bm N}

\def\sA{\scriptscriptstyle A}
\def\sB{\scriptscriptstyle B}
\def\sE{\scriptscriptstyle E}
\def\sP{\scriptscriptstyle P}
\def\sAB{\scriptscriptstyle AB}

\begin{document}


\title{Time transfer functions in Schwarzschild-like metrics in the weak-field \\limit: A unified description of Shapiro and lensing effects}

\author{B. Linet}
\email{Bernard.Linet@lmpt.univ-tours.fr}
\affiliation{Laboratoire de Math\'ematiques et Physique Th\'eorique, CNRS/UMR 7350,\\F\'ed\'eration Denis Poisson, Universit\'e Fran\c{c}ois Rabelais, F-37200 Tours, France}


\author{P. Teyssandier}
\email{Pierre.Teyssandier@obspm.fr}
\affiliation{SYRTE, Observatoire de Paris, PSL Research University, CNRS, Sorbonne Universit\'es, \\
 UPMC Univ. Paris 06, LNE, 61 avenue de l'Observatoire, F-75014 Paris, France
}


\date{\today}

\begin{abstract}
We present a complete analysis of the light rays within the linearized, weak-field approximation of a Schwarzschild-like metric describing the gravitational field of an isolated, spherically symmetric body. We prove in this context the existence of two time transfer functions and we obtain these functions in an exact closed-form. We are led to distinguish two regimes. In the first regime, the two time transfer functions correspond to rays which are confined in regions of spacetime where the weak-field approximation is valid. Such a regime occurs in gravitational lensing configurations with double images of a given source. We find the general expressions of the angular separation and the difference in light travel time between the two images. In the second regime, there exists only one time transfer function corresponding to a light ray remaining in a region of weak field. Performing a Taylor expansion of this function with respect to the gravitational constant, we obtain the Shapiro time delay completed by a series of so-called ``enhanced terms." The enhanced terms beyond the third order are new.  
\begin{description}
\item[PACS numbers]
04.25.-g, 04.80.Cc, 95.10.Jk, 95.30.Sf
\end{description}
\end{abstract}

$ $

\maketitle


\section{Introduction} \label{sec:intro}

The notion of time transfer function is of crucial interest for modeling the geometric optical effects in the metric theories of gravity. In order to precise what we mean by a time transfer function in a given spacetime $V_4$, consider a light ray $\Gamma$ propagating in a region of $V_4$ covered by a single coordinate system $x^{\alpha}=(x^0, \bx)$, with $x^0=ct$ and $\bx=(x^i), i=1, 2, 3$. Let $(x^0_{\sA}, \bx_{\sA})$ be the point-event where $\Gamma$ is emitted and $(x^0_{\sB}, \bx_{\sB})$ the point-event where it is observed. The light travel time $(x^0_{\sB}-x^0_{\sA})/c$ is a function of $x^0_{\sA}, \bx_{\sA}$ and $\bx_{\sB}$ or a function of $ \bx_{\sA}$, $x^0_{\sB},$ and $\bx_{\sB}$, each of these functions depending on the ray $\Gamma$. So we may write
\be
x^0_{\sB}-x^0_{\sA}=c{\cal T}_{e,\scriptscriptstyle\Gamma}(x^0_{\sA}, \bx_{\sA}, \bx_{\sB})=c{\cal T}_{r,\scriptscriptstyle\Gamma}(\bx_{\sA}, x^0_{\sB}, \bx_{\sB}). \label{TTFe}
\ee

Following a terminology introduced in \cite{Leponcin:2004}, ${\cal T}_{e,\scriptscriptstyle\Gamma}$ (resp. ${\cal T}_{r,\scriptscriptstyle\Gamma}$) is called the (coordinate) emission (resp. reception) time transfer function associated with the light ray $\Gamma$. Of course, ${\cal T}_{r,\scriptscriptstyle\Gamma}$ can be in principle computed if ${\cal T}_{e,\scriptscriptstyle\Gamma}$ is known, and vice versa. It must be noted that for any stationary metric written in adapted coordinates, the time transfer functions ${\cal T}_{e,\scriptscriptstyle\Gamma}$ and ${\cal T}_{r,\scriptscriptstyle\Gamma}$ are one and the same function ${\cal T}_{\scriptscriptstyle\Gamma}$ which depends only upon $\bx_{\sA}$ and $\bx_{\sB}$. 

Knowing a time transfer function associated with a light ray enables one to model the range and Doppler observables of space missions (see, e.g., \cite{Ashby:2010,Hees1:2014,Hees2:2014} and Refs. therein). It also allows the precise modeling of modern astrometry, since the propagation direction of the light ray $\Gamma$ can be determined from the first derivatives of ${\cal T}_{e,\scriptscriptstyle\Gamma}$ or ${\cal T}_{r,\scriptscriptstyle\Gamma}$ at the point of observation \cite{Leponcin:2004,Teyssandier:2008,Bertone:2014}. However, in spite of its practical interest, the explicit determination of the whole set of possible time transfer functions in a given spacetime remains an unsolved problem. Even for the Schwarzschild metric in which the exact solutions to the geodesic equations are known (see, e.g., \cite{Chandrasekhar:1983} and Refs. therein), the null geodesics passing through two given points are only determined in an implicit, very hard to work manner \cite{Cadez:2005}. 

Nevertheless, a procedure enabling one to determine a particular type of time transfer functions has been obtained in \cite{Teyssandier:2008} for any metric which may be expanded in a power series of the gravitational constant $G$ as follows:
\be 
g_{\mu\nu}(x^{\alpha},G)=\eta_{\mu\nu}+\sum_{n=1}^{\infty}G^ng_{\mu\nu}^{(n)}(x^{\alpha}), \label{expg}
\ee
where $\eta_{\mu\nu}$ is the Minkowski metric. This procedure is based on the assumption that there exists a null geodesic linking the emitter and the receiver such that the associated reception time transfer function, say ${\cal T}_{r}$, is given by an expansion having the form
\begin{eqnarray} 
c{\cal T}_{r}(\bx_{\sA}, x^0_{\sB}, \bx_{\sB})&=&\vert\bx_{\sB}-\bx_{\sA}\vert \nonumber \\
&&+c\sum_{n=1}^{\infty}{\cal T}^{(n)}_{r}(\bx_{\sA}, x^0_{\sB}, \bx_{\sB}), \label{expTg}
\end{eqnarray}
where ${\cal T}^{(n)}_{r}$ stands for a term of order $G^n$. One of us has proposed to call such a null geodesic a quasi-Minkowskian light ray (see \cite{Teyssandier:2012}). It may be shown that each ${\cal T}^{(n)}_{r}$ can be expressed by integrals involving only the terms of order 1,..., $n-1$ taken along a zeroth-order null straight line passing through $\bx_{\sA}$ and $\bx_{\sB}$. This iterative procedure implies the uniqueness of the quasi-Minkowskian light ray joining $\bx_{\sA}$, $\bx_{\sB}$ for $x^0_{\sB}$ given. It must be pointed out, however, that the existence of such a ray is not ensured, as we shall see later.

A lot of studies at the first order in $G$ have been devoted to the quasi-Minkowskian light rays traveling through the field of moving bodies (see, e.g., \cite{Klioner:1992,Kopeikin:1999,Kopeikin:2002,Klioner:2003,Kopeikin:2007,Zschocke:2015}). For the Schwarzschild-like metrics, ${\cal T}^{(1)}$ is given by the well-known Shapiro formula \cite{Shapiro:1964,Will:1993}. In this special case, the higher order functions ${\cal T}^{(n)}$ have been also explicitly determined up to $n=3$: the expression of ${\cal T}^{(2)}$ is henceforth a classical result (see \cite{Leponcin:2004,Teyssandier:2008,Klioner:2010}, and Refs. therein). The expression of ${\cal T}^{(3)}$ has been found in recent works \cite{Linet:2013,Teyssandier:2014}.

Recent works have revealed that the approximation of first order in $G$ is not sufficient for modeling all the solar system experiments. The analysis of the Cassini mission data has shown that the Shapiro formula must be supplemented by higher-order terms called ``enhanced terms" which become significant for light rays almost grazing the Sun \cite{Ashby:2010}. More recently, a so-called ``enhanced 2PN term" has been taken into account for modeling light rays passing near the limb of the Sun \cite{Hees1:2014} or the giant plants of the solar system \cite{Klioner:2010,Teyssandier:2012}. It is worth noticing that an additional $G^2$-term was already introduced in \cite{Moyer:2000}. However, these higher-order terms are linked to the linear part of the metric perturbation since they only depend on the post-Newtonian parameter $\gamma$ which characterizes the 1PN curvature of space. So it appears that nonlinear contributions to the light travel time must be taken into account even in the linear, weak-field approximation. This feature shows that the expansion in Eq. (\ref{expTg}) is not legitimate when the emitter and the receiver are located in almost diametrically opposite directions.

A similar problem is facing us when calculating the light deflection angle, as it has been already noticed in \cite{Zschocke:2011}. It is clear that the above-mentioned methods are not well suited to address the problems raised by the gravitational lensing, where multiple images appear. At least one of the two rays is obviously not a quasi-Minkowskian ray. The aim of this paper is to remedy this confusing situation for a Schwarzschild-like metric treated in the limit of the linearized, weak-field approximation. 

The paper is organized as follows. In Sec. \ref{sec:generalities} , we specify the linearized, weak-field metric and give the expression of the single time transfer function in the case of the radial light rays. In Sec. \ref{sec:nrngeod}, we find the exact solution to the null geodesic equations. Each ray is a branch of hyperbola skirting round the origin of the polar coordinates and having this origin as a focus. In Sec. \ref{sec:dlr}, we show that there exist two distinct rays passing through two given points and we determine explicitly their impact parameters. In Sec. \ref{sec:ttf}, we derive the full expression of the two corresponding time transfer functions. In Sec. \ref{sec:Sfltd}, we show that essentially two regimes are recovered from our unifying treatment. In Sec. \ref{sec:dirlp}, we determine the direction of light propagation for each possible ray. In Sec. \ref{sec:awfgl}, we apply our results to the gravitational lensing, which corresponds to the first regime. Analyzing the second regime in Sec. \ref{sec:Stdet}, we get the Shapiro time delay completed by a series of enhanced terms. We present concluding remarks in Sec. \ref{sec:conclu}. 

\vspace{5mm}

\section{Generalities} \label{sec:generalities}

Let us consider a static, spherically symmetric spacetime corresponding to a central mass $M$. Assuming the coordinates  $(x^0, \bx)$ to be quasi-Cartesian isotropic coordinates, the metric may be written in the form
\be
ds^2={\cal A}(r)[(dx^0)^2-{\cal U}(r)\delta_{ij}dx^idx^j], \label{ds2}
\ee
where $r=\vert\bx\vert$. The corresponding spherical coordinates $(r,\vartheta ,\varphi )$ will be also used, so that 
\[
\delta_{ij}dx^idx^j = dr^2+r^2d\vartheta^2+r^2\sin^2\vartheta d\varphi^2.
\]

It is well known that the null geodesics considered as point sets are identical for two conformally related metrics (see, e.g., \cite{Joshi:2007}). Consequently, the light rays will be treated here as null geodesics of the metric conformally related to (\ref{ds2}) defined by 
\be
d\tilde{s}^2=(dx^0)^2-{\cal U}(r)\delta_{ij}dx^idx^j. \label{ds2op}
\ee
This metric just involves one gravitational potential ${\cal U}$. Setting $m=GM/c^2$, it is assumed that this gravitational potential can be represented as a power series in $m/r$ at any point exterior to the central mass, that is 
\be 
{\cal U}(r)=1+2\kappa_1\frac{m}{r}+2\sum_{n=2}^{\infty}\kappa_n\frac{m^n}{r^n}, \label{Ug}
\ee
where the quantities $\kappa_n$ are constants linked to the post-Newtonian parameters involved in the metric (\ref{ds2}). In particular, one has
\be
\kappa_1=1+\gamma. \label{kap1}
\ee
In general relativity, $\kappa_1=2$.

Henceforth, we restrict our attention to the linearized, weak-field approximation. This means that the terms of order $m^2/r^2$ in the metric are neglected, so that Eq. (\ref{ds2op}) reduces to 
\begin{equation} 
d\tilde{s}^2=(dx^0)^2-\left(1+2\kappa_1\frac{m}{r}\right)\delta_{ij}dx^idx^j.
\label{cds2}
\end{equation}

We study the light rays joining an emitter located at point $\bx_{\sA}$ and a receiver located at point $\bx_{\sB}$, both located at a finite distance from the origin of coordinates. These light rays are regarded as {\it exact} solutions to the null geodesics equations of the metric (\ref{cds2}). This means that we do not assume from the beginning that the solutions can be represented as power series in $m$. However, restricting to a linearized metric implies that only the light rays confined in regions of spacetime such that $r\gg m$ can be considered as relevant in the present analysis. In accordance with this restriction, it will be always assumed that $r_{\sA}\gg m$ and $r_{\sB}\gg m$ in what follows, with $r_{\sA}=\vert\bx_{\sA}\vert$ and $r_{\sB}=\vert\bx_{\sB}\vert$. 

When the vectors $\bx_{\sA}$ and $\bx_{\sB}$ are collinear and have the same direction, the time transfer function ${\cal T}_{rad}$ associated with the radial null geodesics passing through $\bx_{\sA}$ and $\bx_{\sB}$ is given by an elementary integration:
\begin{widetext}
\be
c{\cal T}_{rad}(r_{\sA},r_{\sB})=\mbox{sgn}(r_{\sB}-r_{\sA})\Bigg\lbrack \sqrt{r_{\sB}(r_{\sB}+2\kappa_1m)}-\sqrt{r_{\sA}(r_{\sA}+2\kappa_1m)}+2\kappa_1m\ln\left(\frac{\sqrt{r_{\sB}}+\sqrt{r_{\sB}+2\kappa_1m}}{\sqrt{r_{\sA}}+\sqrt{r_{\sA}+2\kappa_1m}}\right)\Bigg\rbrack.
\label{Tr1}
\ee
\end{widetext}

Henceforth, we concentrate our attention exclusively on the nonradial light rays.

\section{Nonradial null geodesics} \label{sec:nrngeod}

Since the metric is spherically symmetric, a nonradial light ray joining $\bx_{\sA}$ and  $\bx_{\sB}$ is confined to the plane passing through $\bx_{\sA}$, $\bx_{\sB}$ and the origin $O$ of spatial coordinates. We choose the coordinate system so that $\vartheta=\pi/2$ for this plane. Two constants of the motion can then be formed:
\begin{eqnarray}
&&\frac{dx^0}{d\lambda}=E, \label{E}  \\
&&r^2\left(1+2\kappa_1\frac{m}{r}\right)\frac{d\varphi}{d\lambda}=J, \label{J}
\end{eqnarray} 
where $\lambda$ is an arbitrary affine parameter of the solution. It is easily shown that $d\tilde{s}^2=0$ along a null geodesic implies $E\neq0$. As a consequence, we put
\be \label{b}
b=\frac{J}{E}.
\ee

Clearly, we have $b\neq 0$ for any nonradial null geodesic. In what follows, $b$ is an algebraic quantity. The absolute value $\vert b\vert$ may be interpreted as the impact parameter of the light ray \cite{Chandrasekhar:1983}.  By convention, we suppose that $E>0$, which means that the affine parameter $\lambda$ is increasing with time. So $b$ has the sign of $J$. As a consequence, $b>0$ (resp. $b<0$) when $\varphi$ is an increasing (resp. a decreasing) function of time.

It follows from the null geodesic equations of the metric (\ref{cds2}) that $\varphi$ and $r$ are linked by the differential equation
\be \label{phir}
\left(\frac{d\varphi}{dr}\right)^2=\frac{b^2}{r^2(r^2+2\kappa_1mr-b^2)}.
\ee
This equation coincides with the Newtonian differential equation of the trajectory of a fictitious massive particle gravitating with an hyperbolic motion in the field of a pointlike mass $\kappa_1 M$ and having the speed $c$ at infinity \cite{Landau:1976}, a feature which straightforwardly  implies that the light deflection predicted by the present approximation is exactly $\kappa_1$ times the Newtonian deflection. Each light ray is therefore a branch of hyperbola of which one of the foci is the origin $O$, this branch skirting round $O$. It follows from classical formulas for a Keplerian trajectory that the length of the semitransverse axis $a$ is given by
\be \label{a}
a=\kappa_1m
\ee 
and the length of its semiconjugate axis coincides with the impact parameter $\vert b\vert$.

Denoting by $\varphi_{\sP}$ the value of $\varphi$ at the pericenter, the polar equation of a light ray solution to Eq. (\ref{phir}) may be written as
\be \label{eqH}
\frac{1}{r}=\frac{1}{p}[1+e\cos(\varphi-\varphi_{\sP})],
\ee
where $p$ is the parameter of the hyperbola and $e$ its eccentricity, with these quantities being given by
\begin{eqnarray} 
p=&&\frac{b^2}{\kappa_1m}, \label{p} \\
e=&&\frac{\vert b\vert}{\kappa_1m}\sqrt{1+\frac{\kappa_1^2m^2}{b^2}}.  \label{e}
\end{eqnarray}

A geometrical reasoning due to van der  Woude \cite{Woude:1922} showed that there exist two and only two solutions to Eq. (\ref{phir}) passing through two given points $\bx_{\sA}$ and $\bx_{\sB}$ and having their concavity oriented toward the origin, provided that $\bx_{\sA}$ and $\bx_{\sB}$ are not aligned with $O$. Consequently, there exist two and only two time transfer functions in the background defined by the optical metric (\ref{cds2}). Of course, the time transfer functions coincide when $\bx_{\sA}$ and $\bx_{\sB}$ are aligned with $O$.  

As long as points $\bx_{\sA}$ and $\bx_{\sB}$ are located at finite distances from $O$, the range of $\varphi$ is limited by the inequalities
\be \label{varp}
-\frac{1}{e}<\cos(\varphi-\varphi_{\sP})\leq 1.
\ee

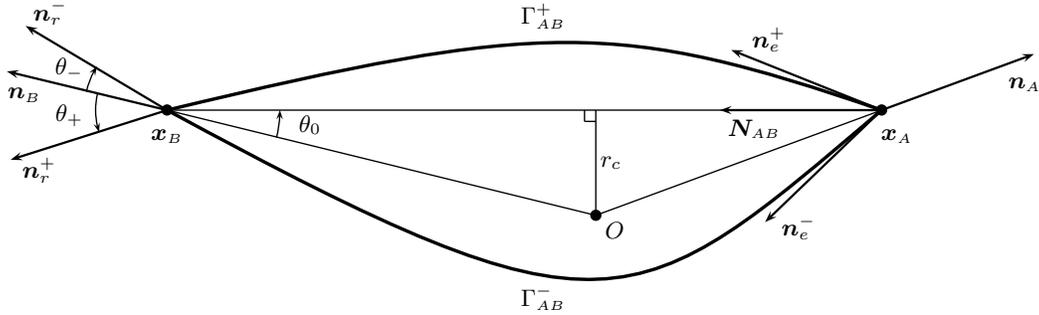
\begin{figure*}
\vspace{2.2cm}
\psscalebox{1.0 1.0} 
{
\begin{pspicture}(0,-0.30)(11.50,0.30)
\rput(1.0,-0.35){$\bx_{\sB}$}
\rput(10.7,-0.35){$\bx_{\sA}$}
\psline[linecolor=black, linewidth=0.018](1.0,0.0)(10.5,0.0)
\psline[linecolor=black, linewidth=0.030]{<-}(8.35,0.0)(10.5,0.0)
\rput(8.8,-0.25){$\bN_{\sAB}$}
\psdots[linecolor=black, dotsize=0.15](1.0,0.0)
\psdots[linecolor=black, dotsize=0.15](10.5,0.0)
\psdots[linecolor=black, dotsize=0.15](6.7,-1.4)
\rput(6.95,-1.6){$O$}
\rput(6.9,-0.7){$r_c$}
\psline[linecolor=black, linewidth=0.018](6.7,-1.4)(6.7,0.0)
\psline[linecolor=black, linewidth=0.018](6.7,-0.15)(6.55,-0.15)
\psline[linecolor=black, linewidth=0.018](6.55,-0.15)(6.55,0.0)
\psline[linecolor=black, linewidth=0.018](6.7,-1.4)(1.0,0.0)
\psline[linecolor=black, linewidth=0.018](6.7,-1.4)(10.5,0.0)
\psbezier[linecolor=black, linewidth=0.045](1.0,0.0)(6.0,1.2)(7.0,1.2)(10.5,0.0)
\psbezier[linecolor=black, linewidth=0.045](1.0,0.0)(6.9,-3.2)(7.2,-2.8)(10.5,0.0)
\rput(6.0,1.25){$\Gamma_{\sAB}^{+}$}
\rput(6.0,-2.50){$\Gamma_{\sAB}^{-}$}
\psline[linecolor=black, linewidth=0.030]{->}(10.5,0.0)(8.5,0.80)
\psline[linecolor=black, linewidth=0.030]{->}(10.5,0.0)(8.95,-1.49)
\psline[linecolor=black, linewidth=0.030]{->}(10.5,0.0)(12.5256,0.7456)
\psline[linecolor=black, linewidth=0.030]{->}(1.0,0.0)(-1.08,-0.66)
\psline[linecolor=black, linewidth=0.030]{->}(1.0,0.0)(-0.88,1.12)
\psline[linecolor=black, linewidth=0.030]{->}(1.0,0.0)(-1.130,0.515)
\rput(9.4,-1.56){$\bn_{e}^{-}$}
\rput(9.0,0.94){$\bn_{e}^{+}$}
\rput(12.4,0.34){$\bn_{\sA}$}
\rput(-0.55,1.3){$\bn_{r}^{-}$}
\rput(-0.90,0.2){$\bn_{\sB}^{}$}
\rput(-0.7,-0.8){$\bn_{r}^{+}$}
\rput(-0.30,0.55){$\theta_{-}$}
\rput(-0.30,-0.07){$\theta_{+}$}
\rput(2.9,-0.2){$\theta_{0}$}
\psarc[linecolor=black, linewidth=0.018]{->}(1.0,0.0){1.5}{-14.0}{0.0}
\psarc[linecolor=black, linewidth=0.018]{->}(1.0,0.0){0.95}{167.0}{198.0}
\psarc[linecolor=black, linewidth=0.018]{<-}(1.0,0.0){1.1}{148.0}{167.0}
\end{pspicture}
}
\vspace{2.6cm}
\caption{The two possible light rays joining two given points within the linearized, weak-field approximation.} 
\label{fig1}
\vspace{1.0cm}
\end{figure*}

In order to determine the time transfer functions, we may assume that $\varphi_{\sA}$ and $\varphi_{\sB}$ are given in such a way that inequalities as follows
\be \label{phAB}
0<\varphi_{\sB}-\varphi_{\sA}<\pi
\ee
hold. We denote by ${\Gamma}_{\sAB}^{+}$ the trajectory of light joining $\bx_{\sA}$ and $\bx_{\sB}$ along which $\varphi$ continuously increases from $\varphi_{\sA}$ to $\varphi_{\sB}$ (see Fig. \ref{fig1}). With our convention that $E>0$, it follows from (\ref{J}) and (\ref{b}) that the parameter $b$ has a value $b_{+}>0$.

There exists also a light ray along which $\varphi$ continuously decreases from $\varphi_{\sA}$ to $\overline{\varphi}_{\sB}$, with $\overline{\varphi}_{\sB}$ being defined by
\be \label{phABb}
\overline{\varphi}_{\sB}=\varphi_{\sB}-2\pi.
\ee     
The trajectory of this light ray will be denoted by ${\Gamma}_{\sAB}^{-}$. With our convention on the sign of $E$, we have $b_{-}<0$.

The time transfer functions corresponding to ${\Gamma}_{\sAB}^{+}$ and to ${\Gamma}_{\sAB}^{-}$ will be denoted by ${\cal T}_{+}$ and ${\cal T}_{-}$, respectively. As we shall see below, the explicit expression of these functions may be inferred from the expressions of $b_{+}$ and $b_{-}$ as functions of $r_{\sA}, r_{\sB}$ and $\varphi_{\sB}-\varphi_{\sA}$. The analytical determination of the quantities $b_{+}$ and $b_{-}$ is the subject of the following section.

\section{Determination of the light rays passing by two points} \label{sec:dlr}

It is straightforwardly inferred from (\ref{eqH}) that $e$, $p$ and $\varphi_{\sP}$ must satisfy two relations as follows:
\begin{subequations}
\label{epP}
\begin{eqnarray} 
\frac{p}{e r_{\sA}}-\frac{1}{e}=\cos(\varphi_{\sA}-\varphi_{\sP}), \label{epPa} \\ 
\frac{p}{e r_{\sB}}-\frac{1}{e}=\cos(\varphi_{\sB}-\varphi_{\sP}). \label{epPb}
\end{eqnarray} 
\end{subequations}
It is possible to eliminate the unknown quantity $\varphi_{\sP}$. Indeed, taking Eqs. (\ref{epP}) into account, we can form the system of equations

\begin{subequations}
\label{epPsc}
\begin{eqnarray}
&&\left(\frac{p}{e r_{\sB}}-\frac{1}{e}\right)\cos\varphi_{\sA}-\left(\frac{p}{e r_{\sA}}-\frac{1}{e}\right)\cos\varphi_{\sB} \nonumber \\
&&\qquad=\sin(\varphi_{\sB}-\varphi_{\sA})\sin\varphi_{\sP}, \label{epPs} \\
&&\left(\frac{p}{e r_{\sA}}-\frac{1}{e}\right)\sin\varphi_{\sB}-\left(\frac{p}{e r_{\sB}}-\frac{1}{e}\right)\sin\varphi_{\sA}\nonumber \\
&&\qquad=\sin(\varphi_{\sB}-\varphi_{\sA})\cos\varphi_{\sP}. \label{epPc}
\end{eqnarray}
\end{subequations}

Squaring (\ref{epPs}) and (\ref{epPc}), and then adding the relations thus obtained, we get an equation independent of $\varphi_{\sP}$ in which we can substitute for $p$ and $e$ from (\ref{p}) and (\ref{e}), respectively. Putting
\be \label{mu}
\mu=\cos(\varphi_{\sB}-\varphi_{\sA}), 
\ee
it is easily seen that this equation may be written in the form 
\begin{eqnarray} 
b^4&&-\frac{r_{\sA}^2r_{\sB}^2}{R_{\sAB}^2}(1-\mu)\left[1+\mu+\frac{2\kappa_1m(r_{\sA}+r_{\sB})}{r_{\sA}r_{\sB}}\right]b^2\nonumber \\
&&+\kappa_1^2m^2\frac{r_{\sA}^2r_{\sB}^2}{R_{\sAB}^2}(1-\mu)^2=0, \label{eqb2}
\end{eqnarray}
where $R_{\sAB}$ is the usual Euclidean distance between $\bx_{\sA}$ and $\bx_{\sB}$ given by
\be \label{RAB}
R_{\sAB}=\vert \bx_{\sB}-\bx_{\sA}\vert=\sqrt{r_{\sA}^2+r_{\sB}^2-2\mu r_{\sA}r_{\sB}}.
\ee

Considered as an equation for $b^2$, Eq. (\ref{eqb2}) has two real roots $b^2_{+}$ and $b^2_{-}$. 

When $\mu=-1$, these roots coincide. Their shared value is then 
\be \label{cr}
b^2_{+}=b^2_{-}=2\kappa_1m\frac{r_{\sA}r_{\sB}}{r_{\sA}+r_{\sB}}.
\ee

When $-1<\mu<1$, the roots of Eq. (\ref{eqb2}) are given by
\begin{eqnarray} 
b^2_{\pm}&=&\frac{r^2_{\sA}r^2_{\sB}(1-\mu^2)}{2R^2_{\sAB}}\Bigg\lbrack 1+\frac{2\kappa_1m(r_{\sA}+r_{\sB})}{r_{\sA}r_{\sB}(1+\mu)}\nonumber \\
&&\pm\sqrt{1+\frac{4\kappa_1m(r_{\sA}+r_{\sB})}{r_{\sA}r_{\sB}(1+\mu)}+\frac{8\kappa_1^2m^2}{r_{\sA}r_{\sB}(1+\mu)}}\Bigg\rbrack .\label{b2pm}
\end{eqnarray}

Both $b_{+}^2$ and $b_{-}^2$ are positive. This property is a direct consequence of the expression of $b_{+}^2$ and of the following relation, inferred from (\ref{eqb2}):
\be \label{bpbm2}
b^2_{+}b^2_{-}=\kappa_1^2m^2\frac{r_{\sA}^2r_{\sB}^2 (1-\mu)^2}{R_{\sAB}^2}.
\ee

Of course, $b_{+}$ and $b_{-}$ could be determined by taking the square root of the right-hand side of Eq. (\ref{b2pm}). However, the expression of these quantities can be notably simplified. Equation (\ref{eqb2}) yields
\be \label{bpabm}
b_{+}^2+b_{-}^2=\frac{r_{\sA}^2r_{\sB}^2(1-\mu)}{R_{\sAB}^2}\left[1+\mu+\frac{2\kappa_1m(r_{\sA}+r_{\sB})}{r_{\sA}r_{\sB}}\right].
\ee
Furthermore, Eq. (\ref{bpbm2}) implies
\be \label{bpbm}
b_{+}b_{-}=-\kappa_1m\frac{r_{\sA}r_{\sB}}{R_{\sAB}}(1-\mu),
\ee
the minus sign in the right-hand side being due to the fact that $b_{+}$ and $b_{-}$ have unlike signs. Then, calculating $(b_{+}+b_{-})^2$ and $(b_{+}-b_{-})^2$ from Eqs. (\ref{bpabm}) and (\ref{bpbm}) straightforwardly lead to expressions as follow for the constants of the motion $b_{+}$ and $b_{-}$:
\begin{widetext}
\begin{equation} 
b_{\pm}=\frac{1}{2}\frac{r_{\sA}r_{\sB}\sqrt{1-\mu}}{R_{\sAB}}\left\lbrace\sqrt{1+\mu+\frac{2\kappa_1m(r_{\sA}+r_{\sB}-R_{\sAB})}{r_{\sA}r_{\sB}}}\pm\sqrt{1+\mu+\frac{2\kappa_1m(r_{\sA}+r_{\sB}+R_{\sAB})}{r_{\sA}r_{\sB}}}\right\rbrace. \label{bpm}
\end{equation}
\end{widetext}

Knowing $b_{+}$ and $b_{-}$, the parameter $p$ and the eccentricity $e$ of the hyperbolas $\Gamma^{+}_{\sAB}$ and $\Gamma^{-}_{\sAB}$ could be inferred from Eqs. (\ref{p}) and (\ref{e}), respectively. To complete the determination of the rays, we would still have to deduce the corresponding values of $\varphi_{\sP}$ from Eqs. (\ref{epP}).

\section{Time transfer functions} \label{sec:ttf}

It follows from Eq. (43) given in \cite{Linet:2013} that the time transfer functions ${\cal T}_{+}$ and ${\cal T}_{-}$ are linked to $b_{+}$ and $b_{-}$ by the differential equations
\be \label{bTmu}
b_{\pm}=-c\sqrt{1-\mu^2}\frac{\partial {\cal T}_{\pm}}{\partial\mu},
\ee 
where $b_{\pm}$ and ${\cal T}_{\pm}$ are regarded as functions of $r_{\sA}, r_{\sB}$ and $\mu$. Integrating Eq. (\ref{bTmu}) leads to  
\begin{eqnarray} 
c{\cal T}_{\pm}(r_{\sA}, r_{\sB},\mu)&=&-\int_{\pm 1}^{\mu}\frac{b_{\pm}(r_{\sA}, r_{\sB},\xi)}{\sqrt{1-\xi^2}}d\xi\nonumber \\
&&+c{\cal T}_{\pm}(r_{\sA}, r_{\sB}, \pm1). \label{iTpm}
\end{eqnarray}

It is clear that ${\cal T}_{+}(r_{\sA}, r_{\sB}, 1)={\cal T}_{rad}(r_{\sA}, r_{\sB})$, where ${\cal T}_{rad}$ is the radial time transfer function given by (\ref{Tr1}). So we have
\be \label{iTp}
c{\cal T}_{+}(r_{\sA}, r_{\sB},\mu)=\int_{\mu}^{1}\frac{b_{+}(r_{\sA}, r_{\sB},\xi)}{\sqrt{1-\xi^2}}d\xi+c{\cal T}_{rad}(r_{\sA}, r_{\sB}).
\ee

The spherical symmetry of spacetime implies 
\[
c{\cal T}_{-}(r_{\sA}, r_{\sB}, -1)=c{\cal T}_{+}(r_{\sA}, r_{\sB}, -1).
\]
Therefore, Eq. (\ref{iTpm}) for  ${\cal T}_{-}$ may be rewritten as
\begin{eqnarray} \label{iTm}
&&c{\cal T}_{-}(r_{\sA}, r_{\sB}, \mu)=-\int_{-1}^{1}\frac{b_{-}(r_{\sA}, r_{\sB},\xi)}{\sqrt{1-\xi^2}}d\xi\nonumber \\
&&+\int_{\mu}^{1}\frac{b_{-}(r_{\sA}, r_{\sB},\xi)}{\sqrt{1-\xi^2}}d\xi+c{\cal T}_{+}(r_{\sA}, r_{\sB}, -1).
\end{eqnarray}

Inserting Eq. (\ref{bpm}) into Eqs. (\ref{iTp}) and (\ref{iTm}) gives
\begin{subequations}
\label{iTpm1}
\begin{eqnarray}
c{\cal T}_{+}(r_{\sA}, r_{\sB}, \mu)&=&I_{+}(\mu)+I_{-}(\mu)\nonumber \\
&&+c{\cal T}_{rad}(r_{\sA}, r_{\sB}), \label{iTp1} \\
c{\cal T}_{-}(r_{\sA}, r_{\sB}, \mu)&=&I_{+}(-1)-I_{-}(-1)
+I_{-}(\mu)\nonumber \\
&&-I_{+}(\mu)+c{\cal T}_{+}(r_{\sA}, r_{\sB}, -1), \label{iTm1}
\end{eqnarray}
\end{subequations}
where $I_{+}$ and $I_{-}$ are defined as 
\be \label{Ipm}
I_{\pm}(\mu)=\frac{r_{\sA} r_{\sB}}{2}\int_{\mu}^{1}\frac{1}{R(\xi)}\sqrt{1+\frac{2\kappa_1m[r_{\sA}+r_{\sB}\pm R(\xi)]}{r_{\sA}r_{\sB}(1+\xi)}}d\xi,
\ee
\begin{widetext}
with $R(\xi)=\sqrt{r_{\sA}^2+r_{\sB}^2-2r_{\sA}r_{\sB}\xi}$. Since $R(\xi)$ is a strictly monotonic function of $\xi$, we can choose it as a new variable for integrating Eqs. (\ref{Ipm}). We get thus
\be \label{Jpm1}
I_{\pm}(\mu)=\frac{1}{2}\int_{\vert r_{\sB}-r_{\sA}\vert}^{R_{\sAB}}
\sqrt{\frac{r_{\sA}+r_{\sB}\mp R+4\kappa_1m}{r_{\sA}+r_{\sB}\mp R}}dR.
\ee
Performing the integration of Eqs. (\ref{Jpm1}), and then taking into account Eq. (\ref{Tr1}) leads to expressions as follows for the time transfer functions:  
\begin{eqnarray}
c{\cal T}_{\pm}(\bx_{\sA}, \bx_{\sB})=&&\frac{1}{2}\left(\sqrt{r_{\sA}+r_{\sB}+R_{\sAB}}\sqrt{r_{\sA}+r_{\sB}+R_{\sAB}+4\kappa_1m}\mp
\sqrt{r_{\sA}+r_{\sB}-R_{\sAB}}\sqrt{r_{\sA}+r_{\sB}-R_{\sAB}+4\kappa_1m}\right)  \nonumber \\
&&+2\kappa_1m\ln\left(\frac{\sqrt{r_{\sA}+r_{\sB}+R_{\sAB}+4\kappa_1m}+\sqrt{r_{\sA}+r_{\sB}+R_{\sAB}}}
{\sqrt{r_{\sA}+r_{\sB}-R_{\sAB}+4\kappa_1m}\pm \sqrt{r_{\sA}+r_{\sB}-R_{\sAB}}}\right).\qquad \qquad \label{TTpm}
\end{eqnarray}
\end{widetext}

We note that Eqs. (\ref{TTpm}) yield expressions of the time transfer functions which are symmetric in $\bx_{\sA}$ and $\bx_{\sB}$, as it could be expected.

When $\bx_{\sA}$ and $\bx_{\sB}$ are located in diametrically opposite directions, i.e. when $\mu=-1$, the expressions of $c{\cal T}_{+}$ and $c{\cal T}_{-}$ coincide and take the simple form
\begin{eqnarray} 
c{\cal T}_{+}&=&c{\cal T}_{-}\nonumber \\
&=&\sqrt{r_{\sA}+r_{\sB}}\sqrt{r_{\sA}+r_{\sB}+2\kappa_1m} \nonumber \\
&&+2\kappa_1m\ln\left(\frac{\sqrt{r_{\sA}+r_{\sB}}+\sqrt{r_{\sA}+r_{\sB}
+2\kappa_1m}}{\sqrt{2\kappa_1m}}\right).\nonumber \\
&&\label{Topp}
\end{eqnarray}
This formula gives the common value of the two time transfer functions for an Einstein ring.

\section{Two standard regimes recovered} \label{sec:Sfltd}

The radial variable takes its minimal value $r_{\sP}$ when the derivative $dr/d\varphi=0$. It is straightforwardly deduced from Eq. (\ref{phir}) that $r_{\sP}^{\pm}$ satisfies the equation
\be \label{erP}
(r_{\sP}^{\pm})^2+2\kappa_1mr_{\sP}^{\pm}-b_{\pm}^2=0.
\ee 
So, the values of $r_{\sP}^{+}$ and $r_{\sP}^{-}$ are given by
\be \label{rP}
r_{\sP}^{\pm}=\sqrt{b_{\pm}^2+\kappa_1^2m^2}-\kappa_1m.
\ee

The calculations performed throughout this work are relevant only if $r\gg m$ at each point of the trajectory of the light ray under examination. Such a condition is obviously met for the Sun or the planets as long as the light ray is not bypassing the central mass. However, in almost all tests of gravitational theories and in gravitational lensing configurations, the light rays are skirting around the central body, and then the condition
\be \label{rPgm}
r_{\sP}^{\pm}\gg m
\ee
must hold. Taking Eq. (\ref{rP}) into account, this means that we must have
\be \label{bgm}
\vert b_{\pm}\vert\gg m.
\ee
So, we have to study the behavior of $b_{\pm}$ and $r_{\sP}^{\pm}$ when $\mu$ is close to $-1$. 

It is easily seen that if $\mu \rightarrow -1$, the derivative of $b^2_{\pm}$ with respect to $\mu$ has an asymptotic behavior given by
\be \label{db2}
\frac{\partial b^2_{\pm}}{\partial\mu}\sim\pm\sqrt{\frac{\kappa_1m}{1+\mu}\left(\frac{r_{\sA}r_{\sB}}{r_{\sA}+r_{\sB}}\right)^{3}\left(1+\frac{2\kappa_1m}{r_{\sA}+r_{\sB}}\right)}.
\ee
It follows from Eq. (\ref{db2}) that $b_{+}^2$ is an increasing function of $\mu$ when $\mu$ is sufficiently close to $-1$. Therefore, $b_{+}^2$ reaches a local lower bound when $\mu=-1$, the value of which is given by Eq. (\ref{cr}). Taking into account Eq. (\ref{rP}), it may be inferred that $r_{\sP}^{+}$ reaches a minimum when $\mu=-1$, namely
\be \label{Ppp}
(r_{\sP}^{+})_{min}=\sqrt{\frac{2\kappa_1mr_{\sA}r_{\sB}}{r_{\sA}+r_{\sB}}+\kappa_1^2m^2}-\kappa_1m.
\ee

Since it is assumed that $r_{\sA}\gg m$ and $r_{\sB}\gg m$, it follows from Eq. (\ref{Ppp}) that $(r_{\sP}^{+})_{min}\gg m$. This means that the ray $\Gamma_{\sAB}^{+}$ is correctly described by the present formalism in the configurations of conjunction or quasiconjunction, i.e. when $\mu$ is in the vicinity of $-1$. This conclusion can be extended to the other values of $\mu$, in spite of the fact that $b_{+}$ is not a monotonic function of $\mu$.

Equation (\ref{db2}) shows that $\vert b_{-}\vert$ is a decreasing function when $\mu \rightarrow -1$. In fact, it may be seen that this property is valid on the range $-1\leq\mu <1$ since Eq. (\ref{bpm}) for $b_{-}$ may be rewritten in the form
\begin{widetext}
\be \label{bm}
b_{-}=-\frac{2\kappa_1m\sqrt{1-\mu}}{\sqrt{1+\mu+\displaystyle \frac{2\kappa_1m(r_{\sA}+r_{\sB}+R_{\sAB})}{r_{\sA}r_{\sB}}}+\sqrt{1+\mu+\displaystyle \frac{2\kappa_1m(r_{\sA}+r_{\sB}-R_{\sAB})}{r_{\sA}r_{\sB}}}}.
\ee
\end{widetext}
Indeed, the numerator $\sqrt{1-\mu}$ in the right-hand side of Eq. (\ref{bm})  is a decreasing function and the denominator an increasing function. So, $r_{\sP}^{-}$ reaches its upper bound $(r_{\sP}^{-})_{max}$ when $\mu=-1$. This upper bound is equal to $(r_{\sP}^{+})_{min}$. Hence 
\be \label{Ppm}
(r_{\sP}^{-})_{max}=\sqrt{\frac{2\kappa_1mr_{\sA}r_{\sB}}{r_{\sA}+r_{\sB}}+\kappa_1^2m^2}-\kappa_1m.
\ee

We are led to distinguish two regimes for the rays $\Gamma^{-}_{\sAB}$.

\subsection{First regime}
The value $\mu=-1$ corresponds to an Einstein ring. We infer from Eq. (\ref{cr}) that $\vert b_{-}\vert \gg m$, and consequently $(r_{\sP}^{-})_{max} \gg m$ hold in this configuration. When $\mu$ is very close to $-1$ or varies so that to have
\be 
1+\mu \propto 1+\mu_{E}, \label{mmE}
\ee
where $\mu_{E}$ is a typical value defined by
\be  
1+\mu_{E}=\frac{\kappa_1m(r_{\sA}+r_{\sB})}{r_{\sA}r_{\sB}},  \label{muE}
\ee
the order of magnitude of $\vert b_{-}\vert$ is given by 
\be \label{bma}
\vert b_{-}\vert \propto \sqrt{\frac{\kappa_1mr_{\sA}r_{\sB}}{r_{\sA}+r_{\sB}}}.
\ee
Consequently, the condition $r_{\sP}^{-}\gg m$ is met, which implies that the light ray $\Gamma_{\sAB}^{-}$ is entirely located in the region where the weak field approximation is valid, as it is also the case for the ray $\Gamma_{\sAB}^{+}$ (see above). We can conclude that this regime corresponds to a configuration of gravitational lensing in a weak field, with two images of the same point-mass object. We will show in Sec. \ref{sec:awfgl} that the well-known formulas of the weak gravitational lensing may be recovered from Eqs. (\ref{TTpm}).

\subsection{Second regime} 
For the values of $\mu$ such that
\be \label{mup1b}
2>1+\mu\gg 1+\mu_{E},
\ee
Eq. (\ref{bm}) implies
\be \label{bmb}
\vert b_{-}\vert \approx \kappa_1m\sqrt{\frac{1-\mu}{1+\mu}}.
\ee
The magnitude of $\vert b_{-}\vert$ is then of the order of $\kappa_1m$. We have therefore $r_{\sP}^{-}\sim \kappa_1m$, a value which is not compatible with the weak-field approximation. So, only the ray $\Gamma_{\sAB}^{+}$ can be treated with our method when the condition (\ref{mup1b}) is met. Configurations of this kind occur in the experiments currently performed in the solar system. We are then facing cases where only the time transfer function ${\cal T}_{+}$ has to be considered. We will see in Sec. \ref{sec:Stdet} how the Shapiro time delay formula supplemented by enhanced terms of any order is recovered from the expression of $c{\cal T}_{+} $ given by Eq. (\ref{TTpm}).

\section{Direction of light propagation} \label{sec:dirlp}

Let $x^{\alpha}=x_{\scriptscriptstyle \Gamma}^{\alpha}(\zeta)$ be a system of parametric equations describing a null geodesic $\Gamma$ joining $\bx_{\sA}$ and $\bx_{\sB}$, $\zeta$ being an arbitrarily chosen parameter. The direction of propagation at any point $x$ of $\Gamma$ is fully determined by the {\it direction triple} defined as \cite{footnote1}
\begin{equation} \label{tl}
\widehat{\bm k}=\left(k_1 /k_0, k_2 /k_0, k_3 /k_0\right),  
\end{equation}
where $k_0$ and $k_i$ are the covariant components of the vector tangent to the ray at $x$, i.e. the quantities $k_{\alpha}=g_{\alpha\beta}(x_{\scriptscriptstyle \Gamma}^{\rho}(\zeta))dx_{\scriptscriptstyle \Gamma}^{\beta}(\zeta)/d\zeta$. This triple does not depend on the choice of the parameter along $\Gamma$.

In the optical metric (\ref{cds2}), the vector $\widehat{\bm k}$ is collinear to the 3-vector tangent to the photon trajectory since $\widehat{\bm k}=-(1+2\kappa_1m/r)d\bm x_{\scriptscriptstyle \Gamma}/dx^0$. It follows from this equation and from the property of  $k_{\alpha}$ to be a null vector that the Euclidean norm of $\widehat{\bm k}$ at any point $x$ of $\Gamma$ is given by
\be
\vert\widehat{\bm k}\vert=\sqrt{1+2\kappa_1\frac{m}{r}}. \label{nlh}
\ee
The propagation direction of light as seen by a static observer staying at $x$ is therefore characterized by the unit {\it direction vector} $\bn$ defined as
\be
\bn=-\frac{\widehat{\bm k}}{\sqrt{1+\displaystyle 2\kappa_1\frac{m}{r}}}. \label{bnh}
\ee

The expressions of the direction vector at points $\bm x_{\scriptscriptstyle A}$ and $\bm x_{\scriptscriptstyle B}$ will be denoted by $\bn_{\,e}(\bm x_{\scriptscriptstyle A},\bm x_{\scriptscriptstyle B})$ and $\bn_{\,r}(\bm x_{\scriptscriptstyle A},\bm x_{\scriptscriptstyle B})$, respectively. These vectors can be derived from the relations 
\begin{subequations}
\label{wl0}
\begin{eqnarray} 
\widehat{\bm k}_{\,e}(\bm x_{\scriptscriptstyle A},\bm x_{\scriptscriptstyle B})=&&c\bm \nabla_{\bx_{\sA}} {\cal T}_{\scriptscriptstyle\Gamma}(\bx_{\sA}, \bx_{\sB}), \label{wlA0} \\ 
\widehat{\bm k}_{\,r}(\bm x_{\scriptscriptstyle A},\bm x_{\scriptscriptstyle B})=&&-c\bm \nabla_{\bx_{\sB}} {\cal T}_{\scriptscriptstyle\Gamma}(\bx_{\sA}, \bx_{\sB}), \label{wlB0}
\end{eqnarray}
\end{subequations}
where ${\cal T}_{\scriptscriptstyle\Gamma}$ is the time transfer function associated to $\Gamma$ and $\bm \nabla_{\bx}$ denotes the gradient with respect to $\bx$ (see \cite{Leponcin:2004} and Refs. therein). 

We denote by $\bn^{+}$ and $\bn^{-}$ the direction vectors relative to $\Gamma_{\sAB}^{+}$ and $\Gamma_{\sAB}^{-}$, respectively. Substituting for $c{\cal T}_{\pm}$ from Eq. (\ref{TTpm}) into Eqs. (\ref{wl0}) yield expressions as follow:
\begin{subequations}
\label{lpm}
\begin{eqnarray} 
\bn_{\,e}^{\pm}(\bm x_{\scriptscriptstyle A},\bm x_{\scriptscriptstyle B})=\frac{1}{2}\frac{(X\pm Y)\bN_{\sAB} 
-(X\mp Y)\bn_{\sA}}{\sqrt{1+2\kappa_1\epsilon_{\sA}}}, \nonumber \\
\label{lepm} \\
\bn_{\,r}^{\pm}(\bm x_{\scriptscriptstyle A},\bm x_{\scriptscriptstyle B})=\frac{1}{2}\frac{(X\pm Y)\bN_{\sAB} 
+(X\mp Y)\bn_{\sB}}{\sqrt{1+2\kappa_1\epsilon_{\sB}}}, \nonumber \\
 \label{lrpm}
\end{eqnarray}
\end{subequations}
where
$X$ and $Y$ are defined by
\begin{eqnarray}
X=\sqrt{1+\frac{4\kappa_1m}{r_{\scriptscriptstyle A}+r_{\scriptscriptstyle B}+ R_{\scriptscriptstyle AB}}}, \label{X}\\
Y=\sqrt{1+\frac{4\kappa_1m}{r_{\scriptscriptstyle A}+r_{\scriptscriptstyle B}- R_{\scriptscriptstyle AB}}}, \label{Y}
\end{eqnarray}
and 
\be
\bN_{\scriptscriptstyle AB}=\frac{\bm x_{\scriptscriptstyle B}-\bm x_{\scriptscriptstyle A}}{\vert\bm x_{\scriptscriptstyle B}-\bm x_{\scriptscriptstyle A}\vert}, \quad \bn_{\scriptscriptstyle A}=\frac{\bm x_{\scriptscriptstyle A}}{r_{\scriptscriptstyle A} },\quad \bn_{\scriptscriptstyle B}=\frac{\bm x_{\scriptscriptstyle B}}{r_{\scriptscriptstyle B}}, \label{Nnn}
\ee
\be
\epsilon_{\sA}=\frac{m}{r_{\sA}}, \quad \epsilon_{\sB}=\frac{m}{r_{\sB}}. \label{epsiB}
\ee

The quantities $X$ and $Y$ may be transformed into
\begin{eqnarray}
X=\sqrt{1+\frac{2\kappa_1m(r_{\scriptscriptstyle A}+r_{\scriptscriptstyle B}- R_{\scriptscriptstyle AB})}{r_{\scriptscriptstyle A}r_{\scriptscriptstyle B}(1+\mu)}}, \label{Xa}\\
Y=\sqrt{1+\frac{2\kappa_1m(r_{\scriptscriptstyle A}+r_{\scriptscriptstyle B}+ R_{\scriptscriptstyle AB})}{r_{\scriptscriptstyle A}r_{\scriptscriptstyle B}(1+\mu)}}. \label{Ya}
\end{eqnarray}
Substituting for $X$ and $Y$ from Eqs. (\ref{Xa}) and (\ref{Ya}) into $X \pm Y$, and then comparing with Eq. (\ref{bpm}) yield 
\be 
X \pm Y=\frac{2b_{\pm}}{r_c}, \label{XpmY}
\ee
where $r_c$ is the Euclidean distance of the straight line passing by $\bx_{\sA}$ and $\bx_{\sB}$ from the origin $O$, namely
\be  
r_c=\frac{r_{\scriptscriptstyle A}r_{\scriptscriptstyle B}\sqrt{1-\mu^2}}{R_{\sAB}}. \label{rc}
\ee
Consequently, Eqs. (\ref{lpm}) may be written in the form 
\begin{subequations}
\label{lpm2}
\begin{eqnarray} 
\bn_{\,e}^{\pm}(\bm x_{\scriptscriptstyle A},\bm x_{\scriptscriptstyle B})=\frac{b_{\pm}\bN_{\sAB} 
-b_{\mp}\bn_{\sA}}{r_c\sqrt{1+2\kappa_1\epsilon_{\sA}}}, 
\label{lepm2} \\
\bn_{\,r}^{\pm}(\bm x_{\scriptscriptstyle A},\bm x_{\scriptscriptstyle B})=\frac{b_{\pm}\bN_{\sAB} 
+b_{\mp}\bn_{\sB}}{r_c\sqrt{1+2\kappa_1\epsilon_{\sB}}}
 \label{lrpm2}.
\end{eqnarray}
\end{subequations}

Let us call $\theta_{+}$ (resp.  $\theta_{-}$) the angle between the propagation direction of a radial light ray going to infinity through $\bx_{\sB}$ and the propagation direction of the light ray $\Gamma^{+}_{\sAB}$ (resp. $\Gamma^{-}_{\sAB}$)  as measured by a static observer staying at point $\bx_{\sB}$ (see Fig. \ref{fig1}). Since we use an isotropic coordinate system adapted to the static character of the metric, it follows from a result shown in \cite{Teyssandier:2012} that $\theta_{+}$ (resp. $\theta_{-}$) is equal  to the Euclidean angle between $\bn_{\sB}$ and the direction vector $\bn_{\,r}^{+}$ (resp. $\bn_{\,r}^{-}$). The angles $\theta_{+}$ and $\theta_{-}$ are regarded as algebraic quantities and their determination is chosen in such a way that $0\leq \vert\theta_{\pm}\vert\leq\pi$, with the convention that $\theta_{+}$ is a positive quantity. Comparing the exterior products $\bn_{\sB}\times\bn_{\,r}^{+}$ and $\bn_{\sB}\times\bn_{\,r}^{-}$ as derived from Eq. (\ref{lrpm2}), it appears that $\theta_{+}$ and $\theta_{-}$ have unlike signs since the signs of $b_{+}$ and $b_{-}$ are unlike. We have therefore an algebraic relation as follows:
\be
\sin\theta_{\pm}=\frac{b_{\pm}}{r_{\sB}\sqrt{1+2\kappa_1\epsilon_{\sB}}}. \label{sthpm}
\ee
To finish determining the values of  $\theta_{+}$ and $\theta_{-}$, it is necessary to know the sign of $\cos\theta_{\pm}$. It follows from Eq. (\ref{lrpm2}) that
\be
\cos\theta_{\pm}=\frac{b_{\pm}\cos\theta_0 
+b_{\mp}}{r_c\sqrt{1+2\kappa_1\epsilon_{\sB}}}, \label{ctpmt0}
\ee
where $\theta_0$ is the angle between $\bn_{\sB}$ and $\bN_{\sAB}$ specified by the relations 
\be
\cos\theta_0=\bn_{\sB}.\bN_{\sAB}=\frac{r_{\sB}-r_{\sA}\mu}{R_{\sAB}},\quad 0\leq\theta_0<\pi .\label{th0}
\ee
Equations (\ref{sthpm})-(\ref{th0}) enable one to express the angles $\theta_{+}$ and $\theta_{-}$ as functions of $r_{\sA}$, $r_{\sB}$ and $\mu$.

Taking into account Eqs. (\ref{bpabm})-(\ref{bpbm}), and then noting that
\be  
\sin\theta_0=\vert\bn_{\sB}\times\bN_{\sAB}\vert=\frac{r_{\sA}\sqrt{1-\mu^2}}{R_{\sAB}}, \label{sith0}
\ee
Eqs. (\ref{sthpm})-(\ref{sith0}) imply
\begin{eqnarray} 
\sin(\theta_{+}+\theta_{-})&=&\sin\theta_0, \nonumber \\
\cos(\theta_{+}+\theta_{-})&=&\cos\theta_0. \nonumber
\end{eqnarray}
Hence the remarkable relation between $\theta_{+}$, $\theta_{-}$ and $\theta_0$
\be 
\theta_{+}+\theta_{-} =\theta_0. \label{tpm0}
\ee 

This relation will be very useful for analyzing the lensing configurations, as it will be seen in the next section. It must be emphasized that Eq. (\ref{tpm0}) holds even if the angles $\theta_{+}$ and $\theta_{-}$ are not small.

\vspace{0.4cm}

\section{Application to the weak-field gravitational lensing} \label{sec:awfgl}

The general formulas of the previous section were established without being worried about the physical meaning of the rays joining $\bx_{\sA}$ and $\bx_{\sB}$. So they may be applied to any configuration of gravitational lensing. 

\subsection{Angular separation between two images}

The angular separation between the two images of the emitter as measured by a static observer staying at $\bx_{\sB}$ is the difference $\theta_{+}-\theta_{-}$. The sine of $\theta_{+}-\theta_{-}$ is easily inferred from the expression of the exterior product $\bn_{\,r}^{-}\times\bn_{\,r}^{+}$. Using Eqs. (\ref{lrpm2}), and then taking into account Eqs. (\ref{b2pm}), we get 
\begin{eqnarray}
\sin(\theta_{+}-\theta_{-})&=&\frac{r_{\sA}\sqrt{1-\mu}}{R_{\sAB}(1+2\kappa_1\epsilon_{\sB})}\nonumber \\
&&\times\sqrt{1+\mu+\frac{4\kappa_1m(r_{\sA}+r_{\sB})}{r_{\sA}r_{\sB}}+\frac{8\kappa_1^2m^2}{r_{\sA}r_{\sB}}} \, .\nonumber \\
&& \label{angs}
\end{eqnarray}

Let us apply the previous results to the case of an Einstein ring, which is characterized by $\theta_0=0$, i.e. $\mu=-1$. Denote by $2\theta_{\sE}$ the corresponding angular separation between the two images. We get from Eq. (\ref{angs})
\be 
\sin 2\theta_{E}=\frac{2}{1+2\kappa_1\epsilon_{\sB}}\sqrt{\frac{2\kappa_1\epsilon_{\sB} r_{\sA}}{r_{\sA}+{r_{\sB}}}\left(1+\frac{2\kappa_1 m}{r_{\sA}+r_{\sB}}\right)}. \label{2thE}
\ee
Equation (\ref{2thE}) is equivalent to a relation as follows
\be 
\sin \theta_{E}=\frac{1}{\sqrt{1+2\kappa_1\epsilon_{\sB}}}\sqrt{\frac{2\kappa_1\epsilon_{\sB}r_{\sA}}{r_{\sA}+r_{\sB}}}, \label{thE}
\ee
which is straightforwardly inferred from Eqs. (\ref{cr}) and (\ref{sthpm}) when the equalities $\theta_{+}=-\theta_{-}=\theta_{E}$ are used.

Taking into account Eq. (\ref{sith0}), a short calculation shows that Eq. (\ref{angs}) may be rewritten in the form
\begin{widetext}
\be
\sin(\theta_{+}-\theta_{-})=\sqrt{\frac{\sin^2\theta_{0}}{(1+2\kappa_1\epsilon_{\sB})^2}+\frac{1-\mu}{2}\left[1+\frac{2r_{\sA}r_{\sB}(1+\mu)}{R^2_{\sAB}}\right]\sin^2 2\theta_{E}} \,. \label{angsa}
\ee
\end{widetext}

In the realistic lensing configurations, $1+\mu$ is very close to 0 or varies in such  way that $1+\mu \propto 1+\mu_{E}$, with $\mu_{E}$ defined by Eq. (\ref{muE}). Then $\theta_0$ is small and $2r_{\sA}r_{\sB}(1+\mu)/R_{\sAB}^2\ll 1$. So $\theta_{+}$ and $\theta_{-}$ are small and Eqs. (\ref{tpm0}) and (\ref{angsa}) enable one to recover the well-known expressions of $\theta_{+}$ and $\theta_{-}$ in lensing configurations, namely
\be
\theta_{\pm}\approx\frac{\theta_{0}}{2}\pm\sqrt{\frac{\theta_0^2}{4}+\theta_E^2}. \label{thpthm}
\ee
Accordingly, Eq. (\ref{thE}) reduces to the usual expression of half the angular diameter of the Einstein ring, that is
\be
\theta_{E}\approx\sqrt{\frac{2\kappa_1m r_{\sA}}{(r_{\sA}+{r_{\sB}) r_{\sB}}}},  \label{thEa}
\ee
and Eq. (\ref{sith0}) shows that $\theta_{0}$ is linked to $1+\mu$ by a relation as follows
\be
\theta_{0}\approx\frac{r_{\sA}}{r_{\sA}+r_{\sB}}\sqrt{2(1+\mu)}. \label{th0mu}
\ee

We have therefore $\theta_{0}\propto\theta_{E}$ when $1+\mu\propto1+\mu_{E}$. We recover the crucial role played by the quantity $\mu_{E}$ defined by Eq. (\ref{muE}) in delimiting the lensing regime. 

\subsection{Difference in light travel times}

The difference in light travel times ${\cal T}_{-}-{\cal T}_{+}$ is relevant since this quantity is observable. We obtain from Eqs. (\ref{TTpm})
\begin{widetext}
\begin{eqnarray}
c({\cal T}_{-}-{\cal T}_{+})&=&\sqrt{r_{\sA}+r_{\sB}-R_{\sAB}}\sqrt{r_{\sA}+r_{\sB}-R_{\sAB}+4\kappa_1m}  \nonumber \\
&&+4\kappa_1m\ln\left(\frac{\sqrt{r_{\sA}+r_{\sB}-R_{\sAB}}+\sqrt{r_{\sA}+r_{\sB}-R_{\sAB}+4\kappa_1m}}{2\sqrt{\kappa_1m}}\right).
\label{TD}
\end{eqnarray}
\end{widetext}
Since the argument of the logarithm in the right-hand side of Eq. (\ref{TD}) is always greater than or equal to $1$, ${\cal T}_{-}-{\cal T}_{+}\geq 0$ for any lensing configuration.

An expression of ${\cal T}_{-}-{\cal T}_{+}$ in terms of the observable angles $\theta_{+}$ and $\theta_{-}$ is possible. Taking into account the fact that Eq. (\ref{Y}) implies
\vspace{-3mm}
\be
r_{\sA}+r_{\sB}-R_{\sAB}=\frac{4\kappa_1m}{Y^2-1}, \label{rrmR}
\ee
it may be seen that Eq. (\ref{TD}) reads
\be 
c({\cal T}_{-}-{\cal T}_{+})=\frac{4\kappa_1mY}{Y^2-1}+4\kappa_1m\ln\sqrt{\frac{Y+1}{Y-1}}. \label{TD1}
\ee
It follows from Eqs. (\ref{XpmY}), (\ref{rc}) and (\ref{sthpm}) that
\be
Y=\sqrt{1+2\kappa_1\varepsilon_{\sB}}\,\frac{R_{\sAB}}{r_{\sA}\sqrt{1-\mu^2}}(\sin\theta_{+}-\sin\theta_{-}). \nonumber
\ee
Taking into account Eqs. (\ref{sith0}) and (\ref{tpm0}), this equation may be written as
\be
Y=\sqrt{1+2\kappa_1\varepsilon_{\sB}}\,\frac{\sin\theta_{+}-\sin\theta_{-}}{\sin(\theta_{+}+\theta_{-})}.  \label{Y1}
\ee
Substituting for $Y$ from Eq. (\ref{Y1}) into Eq. (\ref{TD1}), and then noting that $Y>1$ and that $\sin(\theta_{+}+\theta_{-})>0$ follows from Eqs. (\ref{th0}) and (\ref{tpm0}), it is easily checked that the difference in light travel time due to the gravitational lensing may be written in the form
\begin{widetext}
\begin{eqnarray}
c({\cal T}_{-}-{\cal T}_{+})=&&4\kappa_1m\,\frac{\sqrt{1+2\kappa_1\varepsilon_{\sB}}(\sin\theta_{+}-\sin\theta_{-})\sin(\theta_{+}+\theta_{-})}{(1+2\kappa_1\varepsilon_{\sB})(\sin\theta_{+}-\sin\theta_{-})^2-\sin^2(\theta_{+}+\theta_{-})} \nonumber \\
&&+2\kappa_1m\ln\left[\frac{\sqrt{1+2\kappa_1\varepsilon_{\sB}}(\sin\theta_{+}-\sin\theta_{-})+\sin(\theta_{+}+\theta_{-})}{\sqrt{1+2\kappa_1\varepsilon_{\sB}}(\sin\theta_{+}-\sin\theta_{-})-\sin(\theta_{+}+\theta_{-})}\right]. \label{TD2}
\end{eqnarray}
\end{widetext}

As it has been pointed out in the discussion of the apparent angles, the terms of order $\varepsilon_{\sB}$ in Eq. (\ref{TD2}) can be neglected and the angles $\theta_{+}$ and $\theta_{-}$ are very small in astrophysical observations of double images. In practice, Eq. (\ref{TD2}) reduces therefore to
\be
c({\cal T}_{-}-{\cal T}_{+})\approx\kappa_1m\frac{\theta_{+}^2-\theta_{-}^2}{\vert\theta_{+}\theta_{-}\vert}+2\kappa_1m\ln\left\vert\frac{\theta_{+}}{\theta_{-}}\right\vert. \label{TD3}
\ee

In general relativity, Eq. (\ref{TD3}) coincides with the formula (4.78a) in \cite{Schneider:1992} giving the difference in light travel time due to the gravitational lensing when the cosmological redshift is neglected.

\section{Shapiro time delay and enhanced terms} \label{sec:Stdet}

After some calculations, the function $c{\cal T}_{+}$ given by Eq. (\ref{TTpm}) may be rewritten in the form
\begin{eqnarray}
c{\cal T}_{+}&=&R_{\sAB}+\kappa_1m\ln\left(\frac{r_{\sA}+r_{\sB}+R_{\sAB}}{r_{\sA}+r_{\sB}-R_{\sAB}}\right) \nonumber \\
&&+\frac{8\kappa_1^2m^2R_{\sAB}}{r_{\sA}r_{\sB}(1+\mu)}\frac{1}{(1+X)(1+Y)(X+Y)} \nonumber \\
&&+2\kappa_1m\ln\left(\frac{1+X}{1+Y}\right). \label{Tp}
\end{eqnarray}
The two first terms in the right-hand side of Eq. (\ref{Tp}) are the Euclidean distance between $\bx_{\sA}$ and $\bx_{\sB}$ supplemented by the Shapiro time delay \cite{Shapiro:1964,Will:1993}. Substituting Eqs. (\ref{Xa}) and (\ref{Ya}) into Eq. (\ref{Tp}), and then performing a Taylor expansion about $m=0$ yields 
\be
c{\cal T}_{+}=R_{\sAB}+\kappa_1m\ln\left(\frac{r_{\sA}+r_{\sB}+R_{\sAB}}{r_{\sA}+r_{\sB}-R_{\sAB}}\right)+c\sum_{n=2}^{\infty}{\cal T}_{+}^{(n)} , \label{expT}
\ee
where each function ${\cal T}_{+}^{(n)}$ may be written in a form as $\kappa_1^nm^n\Theta^{(n)}_{+}(r_{\sA}, r_{\sB}, \mu)/(1+\mu)^{n-1}$, $\Theta^{(n)}_{+}$ being a regular function when $\mu\rightarrow -1$. Consequently, in this regime, $\Gamma^{+}_{\sAB}$ is a quasi-Minkowskian light ray of the metric (\ref{cds2}).

To determine the values of $\mu$ for which this expansion is convergent, it is convenient to return to Eq. (\ref{iTpm1}). It follows from (\ref{Tr1}) and (\ref{Jpm1}) that ${\cal T}_{rad}$ and $I_{-}(\mu)$ can be developed in power series of $m$ whatever $\mu$. In contrast, it is easily seen that the integrand in Eq. (\ref{Jpm1}) written for $I_{+}(\mu)$ can be expanded in a power series in $m$ for any $R$ in the range $[\vert r_{\sB}-r_{\sA}\vert,R_{\sAB}]$ if and only if $R_{\sAB}\leq r_{\sA}+r_{\sB}-4\kappa_1m$, a condition which is equivalent to
\be 
1+\mu \geq\frac{4\kappa_1m(r_{\sA}+r_{\sB}-2\kappa_1m)}{r_{\sA}r_{\sB}}. \label{can2}
\ee
Consequently, $I_{+}(\mu)$ can be expanded as a power series if the condition (\ref{can2}) is met. This feature and the fact that the quantity $Y$ involved in Eq. (\ref{Tp}) cannot be expanded in a power series in $m$ if $R_{\sAB}>r_{\sA}+r_{\sB}-4\kappa_1m$ enable one to conclude that the expansion in Eq. (\ref{expT}) is convergent if and only if the inequality (\ref{can2}) is satisfied.  

In practice, it can be said that the Taylor series (\ref{expT}) is convergent if $1+\mu \geq 4(1+\mu_E)$, {\it a fortiori} if the inequalities (\ref{mup1b}) are satisfied. For $n=2,\ldots,5$, the functions ${\cal T}_{+}^{(n)}$ are given by 
\begin{eqnarray}
c{\cal T}_{+}^{(2)}(\bx_{\sA}, \bx_{\sB})=&&-\kappa_1^2m^2\frac{R_{\sAB}}{r_{\sA}r_{\sB}(1+\mu)}, \label{Tp2} \\
c{\cal T}_{+}^{(3)}(\bx_{\sA}, \bx_{\sB})=&&\kappa_1^3m^3\frac{R_{\sAB}(r_{\sA}+r_{\sB})}{r_{\sA}^2r_{\sB}^2(1+\mu)^2}, \label{Tp3} \\
&& \nonumber \\
c{\cal T}_{+}^{(4)}(\bx_{\sA}, \bx_{\sB})=&&-\frac{5}{12}\kappa_1^4m^4\frac{R_{\sAB}[3(r_{\sA}+r_{\sB})^2+R_{\sAB}^2]}{r_{\sA}^3r_{\sB}^3(1+\mu)^3}, \nonumber \\
&& \label{Tp4}\\
&& \nonumber \\
c{\cal T}_{+}^{(5)}(\bx_{\sA}, \bx_{\sB})=&&\frac{7}{4}\kappa_1^5m^5\frac{R_{\sAB}(r_{\sA}+r_{\sB})[(r_{\sA}+r_{\sB})^2+R_{\sAB}^2]}{r_{\sA}^4r_{\sB}^4(1+\mu)^4}. \nonumber \\
&&\label{Tp5}
\end{eqnarray} 

Let us compare these results with the known terms in the right-hand side of Eq. (\ref{expTg}) which correspond to the metric defined by Eqs. (\ref{ds2op}) and (\ref{Ug}). We have for ${\cal T}^{(2)}$ (see \cite{Leponcin:2004, Teyssandier:2008,Klioner:2010}) and  ${\cal T}^{(3)}$ (see \cite{Linet:2013,Teyssandier:2014})
\begin{eqnarray}
c{\cal T}^{(2)}( \bx_{\sA}, \bx_{\sB}) &=& -m^2\frac{R_{\sAB}}{r_{\sA}r_{\sB}} \left(\frac{\kappa_1^2}{1+\mu} -\kappa_2\frac{\arccos\mu}{\sqrt{1-\mu^2}}\right), \nonumber \\
&&  \label{Tqm2}  \\
c{\cal T}^{(3)}( \bx_{\sA}, \bx_{\sB})&=&m^3\frac{R_{\sAB}(r_{\sA}+r_{\sB})}{r_{\sA}^2r_{\sB}^2(1+\mu)} \nonumber \\
&&\times \left(\frac{\kappa_1^3}{1+\mu}-\kappa_1\kappa_2\frac{\arccos\mu}{\sqrt{1-\mu^2}}+\kappa_3\right), \nonumber \\
&& \label{Tqm3}
\end{eqnarray}
where $\kappa_2$ and $\kappa_3$ are coefficients appearing in Eq. (\ref{Ug}).

The function ${\cal T}_{+}^{(2)}$ yielded by Eq. (\ref{Tp2}) corresponds to the so-called ``enhanced term" which appears in the right-hand side of Eq. (\ref{Tqm2}) (see \cite{Klioner:2010,Ashby:2010,Teyssandier:2012}). This term is usually regarded as worrying on the grounds of its unboundedness when $1+\mu\rightarrow 0$. It must be emphasized, however, that ${\cal T}_{+}^{(2)}$ is relevant only when the condition in (\ref{can2}) is satisfied. As a consequence, the divergence of ${\cal T}^{(2)}$ when $1+\mu\rightarrow 0$ has no physical meaning, a feature which is confirmed by the property of the function ${\cal T}_{+}$ to be regular on the range $0\leq1+\mu\leq2$. The same analysis holds for ${\cal T}_{+}^{(3)}$, which corresponds to the enhanced term proportional to $\kappa_1^3m^3/(1+\mu)^{2}$ occurring in the term ${\cal T}^{(3)}$ given by Eq. (\ref{Tqm3}). So, it is natural to conjecture that the dominant enhanced term appearing in the perturbation function ${\cal T}^{(n)}$ corresponds to the term ${\cal T}_{+}^{(n)}$ for any $n$.

To finish, let us note that the expansion
\begin{widetext}
\be
\kappa_1m\ln\left(\frac{r_{\sA}+r_{\sB}+R_{\sAB}+\kappa_1m}{r_{\sA}+r_{\sB}-R_{\sAB}+\kappa_1m}\right) =\kappa_1m\ln\left(\frac{r_{\sA}+r_{\sB}+R_{\sAB}}{r_{\sA}+r_{\sB}-R_{\sAB}}\right)+c\left({\cal T}_{+}^{(2)}+\frac{1}{2}{\cal T}_{+}^{(3)}+\frac{1}{5}{\cal T}_{+}^{(4)}+\frac{1}{14}{\cal T}_{+}^{(5)}+\cdots \right) \label{Moy1}
\ee
\end{widetext}
is valid when inequality (\ref{can2}) holds. As a consequence, 
Eq. (\ref{expT}) may be rewritten as an expansion from which
the term $T_{+}^{(2)}$ is removed:
\begin{eqnarray}
c{\cal T}_{+}&=&R_{\sAB}+\kappa_1m\ln\left(\frac{r_{\sA}+r_{\sB}+R_{\sAB}+\kappa_1m}{r_{\sA}+r_{\sB}-R_{\sAB}+\kappa_1m}\right)\nonumber \\
&&+c\left(\frac{1}{2}{\cal T}_{+}^{(3)}+\frac{4}{5}{\cal T}_{+}^{(4)}+\frac{13}{14}{\cal T}_{+}^{(5)}+\cdots\right). \label{Moy2}
\end{eqnarray}
If one neglects the terms of order higher than two, Eq. (\ref{Moy2}) reduces to the expression of the time transfer function introduced by Moyer on the basis of a rather difficult reasoning \cite{Moyer:2000}. So Moyer's formula incorporates the dominant term in Eq. (\ref{Tqm2}). This is the reason for using Moyer's expression instead of Shapiro's formula in the discussions of optical effects in the solar system as long as the third-order term ${\cal T}_{+}^{(3)}$ may be neglected (see, e.g., Refs. \cite{Ashby:2010, Hees1:2014,Hees2:2014,Bertone:2014}).

\section{Conclusion} \label{sec:conclu}

In this paper we have developed a complete description of light rays within the linearized, weak-field approximation of the Schwarzschild-like metrics. Our main result is the explicit computation of the two possible time transfer functions ${\cal T}_{+}$ and ${\cal T}_{-}$ relative to the optical metric (\ref{cds2}). The expressions of these functions yielded by Eq. (\ref{TTpm}) are valid for any points $\bx_{\sA}$ and $\bx_{\sB}$, provided that the corresponding light rays are confined in the region $r\gg m$.

Our analysis leads to distinguish two standard regimes. The first regime corresponds to gravitational lensing configurations in which two images of a given source can be observed. The second regime corresponds to the cases where the light ray associated with ${\cal T}_{-}$ has an impact parameter of the order of $m$, which implies that only the ray corresponding to ${\cal T}_{+}$ has to be kept in our scheme.

We carry out the calculation of the propagation direction of each possible ray. Then, for gravitational lensing configurations, we find the general expressions yielding the angular separation and the difference in light travel time between the two images of a source. When the angles are very small, our formulas reduce to the usual relations for the static, spherically symmetric lensing.

In the second regime, we expand the function ${\cal T}_{+}$ as a Taylor series in powers of $m$. It appears that this series is convergent if and only if the condition in Eq. (\ref{can2}) is met. Explicit calculations of the terms involved in the expansion are carried out up to the fifth order. The first-order term is the Shapiro time delay. A comparison of the second- and third-order terms with Eqs. (\ref{Tqm2}) and (\ref{Tqm3}) strongly suggests that ${\cal T}_{+}^{(n)}$ corresponds to the dominant enhanced term in ${\cal T}^{(n)}$ for any $n$. As a final remark, we point out that the second-order term we find in the expansion of ${\cal T}_{+}$ can be absorbed in the logarithmic function proposed by Moyer, as it is shown by Eq. (\ref{Moy2}). 

\nocite{*}

\bibliography{apssamp}

\begin{thebibliography}{99}

\bibitem{Leponcin:2004} C. Le Poncin-Lafitte, B. Linet and P. Teyssandier, Classical Quantum Gravity {\bf 21}, 4463 (2004).

\bibitem{Ashby:2010} N. Ashby and B. Bertotti, Classical Quantum Gravity {\bf 27}, 145013 (2010).

\bibitem{Hees1:2014} A. Hees, S. Bertone and C. Le Poncin-Lafitte, Phys. Rev. D {\bf 89}, 064045 (2014).

\bibitem{Hees2:2014} A. Hees, S. Bertone and C. Le Poncin-Lafitte, Phys. Rev. D {\bf 90}, 084020 (2014).

\bibitem{Teyssandier:2008} P. Teyssandier and C. Le Poncin-Lafitte, Classical Quantum Gravity {\bf 25}, 145020 (2008).

\bibitem{Bertone:2014} S. Bertone, O. Minazzoli, M. Crosta, C. Le Poncin-Lafitte, A. Vecchiato and M.-C. Angonin, Classical Quantum Gravity {\bf 31}, 015021 (2014).

\bibitem{Chandrasekhar:1983} S. Chandrasekhar, {\it The Mathematical Theory of Black Holes} (Oxford University Press, New York, 1983).

\bibitem{Cadez:2005} A. \v{C}ade\v{z} and U. Kosti\'c, Phys. Rev. D {\bf 72}, 104024 (2005).

\bibitem{Teyssandier:2012} P. Teyssandier, Classical Quantum Gravity {\bf 29}, 245010 (2012).

\bibitem{Klioner:1992}  S. A. Klioner and S. M. Kopeikin, Astron. J. {\bf 104}, 897 (1992).

\bibitem{Kopeikin:1999} S. M. Kopeikin and G. Sch\"{a}fer, Phys. Rev. D {\bf 60}, 124002 (1999).

\bibitem{Kopeikin:2002} S. M. Kopeikin and B. Mashhoon, Phys. Rev. D {\bf 65} 064025 (2002).

\bibitem{Klioner:2003} S. A. Klioner, Astron. J. {\bf 125}, 1580 (2003).

\bibitem{Kopeikin:2007} S. M. Kopeikin and V. Makarov, Phys. Rev. D {\bf 75}, 062002 (2007).

\bibitem{Zschocke:2015} S. Zschocke, Phys. Rev. D {\bf 92}, 063015 (2015).

\bibitem{Shapiro:1964} I. I. Shapiro, Phys. Rev. Lett. {\bf 13}, 789 (1964).

\bibitem{Will:1993} C. M. Will, {\it Theory and Experiment in Gravitational Physics}, 2nd ed. (Cambridge University Press, Cambridge, England, 1993).

\bibitem{Klioner:2010} S. A. Klioner and S. Zschocke, Classical Quantum Gravity {\bf 27}, 075015 (2010).

\bibitem{Linet:2013} B. Linet and P. Teyssandier, Classical Quantum Gravity {\bf 30}, 175008 (2013).

\bibitem{Teyssandier:2014} P. Teyssandier, in {\it Frontiers in Relativistic Celestial Mechanics}, edited by S. M. Kopeikin, (De Gruyter, Berlin, 2014), vol. 2, p. 1 (arXiv:1407.4361).

\bibitem{Moyer:2000} T. D. Moyer, {\it Formulation for Observed and Computed Values of Deep Space Network Data Types for Navigation}, JPL Publication 00-7, edited by J. H. Yuen (Wiley, New York, 2000).

\bibitem{Zschocke:2011} S. Zschocke, Classical Quantum Gravity {\bf 28}, 125016 (2011).

\bibitem{Joshi:2007} P. S. Joshi, {\it Gravitational Collapse and Spacetime Singularities} (Cambridge University Press, Cambridge, 2007).

\bibitem{Landau:1976} L. D. Landau and E. M. Lifshitz, {\it Mechanics}, 3rd ed. (Elsevier, Amsterdam, 1976).

\bibitem{Woude:1922} W. van der Woude, Proc. K. Akad. Amsterdam {\bf 24}, 187 (1922) (the paper can be found at www.dwc.knaw.nl/DL/publications/PU00014793.pdf). Note that the author erroneously indicated a semi-transverse axis equal to twice the correct value. 

\bibitem{Schneider:1992} P. Schneider, J. Ehlers and E. E. Falco, {\it Gravitational Lenses} (Springer-Verlag, Berlin, 1992).

\bibitem{footnote1} The direction triple defined by Eq. (\ref{tl}) is denoted by $\widehat{\underline{{\bm l}}}$ in Refs. \cite{Teyssandier:2012} and \cite{Teyssandier:2014}.

\end{thebibliography}

\end{document}